\documentclass[10pt, conference, compsocconf]{IEEEtran}

\usepackage{times}
\usepackage{subfigure}
\usepackage{amsthm}
\usepackage{times}
\usepackage{multirow}
\usepackage{graphicx}
\usepackage{xspace}
\usepackage{color}
\usepackage{pifont}
\usepackage{eurosym}
\usepackage{listings}
\usepackage{url}
\usepackage{bm}

\DeclareGraphicsExtensions{.ps,.eps,.pdf}
\usepackage{amsmath,amssymb}
\usepackage{dsfont}

\usepackage{triotex}

\usepackage{pgf}
\usepackage{tikz}
\usetikzlibrary{arrows,automata}
\DeclareGraphicsExtensions{.ps,.eps,.pdf}

\newcommand{\U}{\mathbf{U}}
\newcommand{\Snc}{\mathbf{S}}
\newcommand{\T}{\mathbf{T}}
\newcommand{\R}{\mathbf{R}}
\newcommand{\N}{\mathbb{N}}
\newcommand{\Z}{\mathbb{Z}}
\newcommand{\X}{\mathbf{X}}
\newcommand{\Y}{\mathbf{Y}}
\newcommand{\Zed}{\mathbf{Z}}
\newcommand{\iFF}{\Leftrightarrow}

\newtheorem{theorem}{Theorem}

\newcommand{\footnoteremember}[2]{\footnote{#2}\newcounter{#1}\setcounter{#1}{\value{footnote}}}

\begin{document}

\title{SMT-based Verification of LTL Specifications with Integer Constraints and its Application to Runtime Checking of Service Substitutability}

\author{

\IEEEauthorblockN{Marcello M. Bersani}
\IEEEauthorblockA{Politecnico di Milano\\
Milano, Italy\\
bersani@elet.polimi.it}
\and

\IEEEauthorblockN{Luca Cavallaro}
\IEEEauthorblockA{Politecnico di Milano\\
Milano, Italy\\
cavallaro@elet.polimi.it}
\and

\IEEEauthorblockN{Achille Frigeri}
\IEEEauthorblockA{Politecnico di Milano\\
Milano, Italy\\
frigeri@elet.polimi.it}
\and

\IEEEauthorblockN{Matteo Pradella}
\IEEEauthorblockA{CNR IEIIT-MI\\
Milano, Italy\\
pradella@elet.polimi.it}
\and

\IEEEauthorblockN{Matteo Rossi}
\IEEEauthorblockA{Politecnico di Milano\\
Milano, Italy\\
rossi@elet.polimi.it}
}

\maketitle

\begin{abstract}

An important problem that arises during the execution of service-based applications concerns the ability to determine whether a running service can be substituted with one with a different interface, for example if the former is no longer available.
Standard Bounded Model Checking techniques can be used to perform this check,
but they must be able to provide answers very quickly, lest the check hampers
the operativeness of the application, instead of aiding it.
The problem becomes even more complex when \emph{conversational services} are
considered, i.e., services that expose operations that have Input/Output
data dependencies among them.
In this paper we introduce a formal verification technique for an extension of
Linear Temporal Logic that allows users to include in formulae constraints on
integer variables.
This technique applied to the substitutability problem for conversational
services is shown to be considerably faster and with smaller memory footprint
than existing ones.


\end{abstract}

\section{Introduction}\label{sec:intro}

Service Oriented Architectures (SOAs) are a flexible set of design principles that promote interoperability among loosely coupled services that can be used across multiple business domains.
In this context applications are typically composed of services made available by third-party vendors.
This opens new scenarios that are unimaginable in traditional applications.
On the one hand, an organization does not have total control of every part of the application, hence failures and service unavailability should be taken into account at runtime.
On the other hand, during the application execution new services might become available that enable new features or provide equivalent functionalities with better quality.
Therefore the ability to support the evolution of service compositions, for example by allowing applications to substitute existing services with others discovered at runtime, becomes crucial.

Most of the frameworks proposed in recent years for the runtime management of service compositions make the assumption that all semantically equivalent services agree on their interface~\cite{Antonellis06,Verma2005}.
In the practice this assumption turns out to be unfounded.
The picture is further complicated when one considers \emph{conversational services}, i.e., services that expose operations with input/output data dependencies among them.
In fact, in this case the composition must deal with \emph{sequences} of operation invocations, i.e., the \emph{behavior protocol}, instead of single, independent, ones.

\cite{CNP09, seams10Adapt} propose an approach to tackle the substitutability problem, i.e., the problem of deciding when a service can be dynamically substituted by another one discovered at runtime, based on Bounded Model Checking (BMC) techniques. 
Even if the approach proved to be quite effective, the Propositional Satisfiability (SAT) problem on which traditional BMC relies requires to deal with lengthy constraints, which typically limits the efficiency of the analysis phase.
In the setting of the runtime management of service compositions this is not acceptable, as delays incurred when deciding whether services are substitutable or not can hamper the operativeness of the application.

In this paper we introduce a verification technique, based on Satisfiability Modulo Theories (SMT), for an extension of Propositional Linear Temporal Logic with Both past and future operators (PLTLB).
This extension, called CLTLB(DL), allows users to define formulae including Difference Logic (DL) constraints on time-varying integer variables.

Our SMT-based verification technique has two main advantages:
(i) unlike in traditional BMC, arithmetic domains are not approximated by means of a finite representation, which proves to be particularly useful in the service substitutability problem;
(ii) the implemented prototype is shown to be considerably faster and with smaller memory footprint than existing ones based on traditional BMC, due to the conciseness of the problem encoding.

The technique exploits decidable arithmetic theories supported by many SMT solvers \cite{RT06} to natively deal with integer variables (hence, with an infinite domain).
This allows us to decide larger substitutability problems than before, in significantly less time: the response times of our prototype tool make it usable also in a runtime checking setting.

This paper is structured as follows: Section \ref{sec:applications} introduces the issues underlying the runtime checking of service substitutability; 
Sections \ref{sec:CLTL} and \ref{sec:encoding} present, respectively, CLTLB(DL) and its SMT-based encoding for verification purposes; 
Section \ref{sec:casestudy} explains how the approach works on a case study, and Section \ref{sec:experimental} discusses some experimental results; 
Finally, Section \ref{sec:relatedwork} presents some related works.
\section{Substitutability Checking of Conversational Services}
\label{sec:applications}


The approach presented in~\cite{CNP09} enables service substitution through the automatic definition of suitable \emph{mapping scripts}. These map the sequences of operations that the client is assuming to invoke on the \emph{expected service} into the corresponding sequences made available by the \emph{actual service} (i.e., the service that will be actually used).
Mapping scripts are automatically derived given (i) a description of service interfaces in which input and output parameters are associated with each service operation, and (ii) the behavioral protocol associated with each service, described through an automaton.

The mapping between an expected and an actual service assumes that two compatibility relationships have been previously defined.
The first states the \emph{compatibility between states} of two automata.
The second concerns the \emph{compatibility between names and data} associated with some operation $o_{exp} \in O_{exp}$ in the expected service and those associated with some operation $o'_{act} \in O_{act}$ in the actual service.
For the sake of simplicity, here we assume that states and operation names and data are compatible if they are called the same way (more sophisticated compatibility relationships are explored in~\cite{cavallaro-pesos-09}).

Given these definitions, we say that a sequence of operations in the automaton
of the expected service is \emph{substitutable} by another sequence of
operations in the automaton of the actual service if a client designed to use the expected service sequence can use the actual service sequence without noticing the difference.
This can happen when the following conditions hold:
\begin{enumerate}
\item The sequence in the actual service automaton starts and ends in states that are compatible with the initial and final states of the sequence in the expected service automaton.
\item  All data parameters of the operations in the actual service automaton sequence are compatible with those appearing in the expected service automaton sequence.
\end{enumerate}
This substitutability definition allows us to build a reasoning mechanism based on PLTLB that, given an expected service sequence, returns a corresponding actual service sequence.

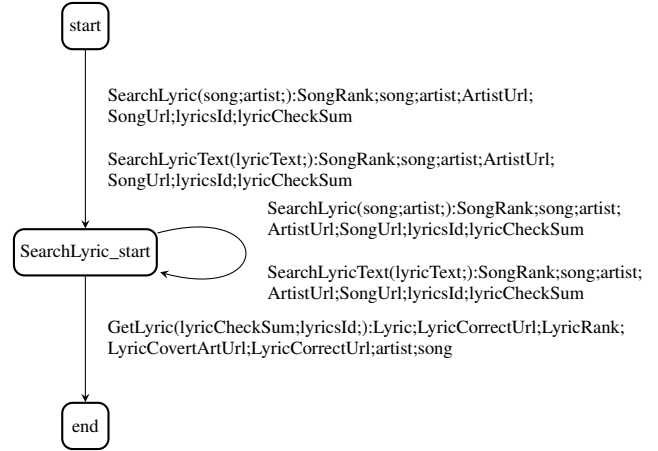
\begin{figure}\centering
\begin{tikzpicture}[node distance=4.3cm,auto,
                    every state/.style={rectangle, text centered, rounded corners, thick, draw=black!100,scale=0.7}, >=stealth]

  \node[state] (start) {start};
  \node[state] (Search) [below of=start] {SearchLyric\_start};
  \node[state] (end) [below of=Search, yshift=1cm] {end};

  \path[->] (start) edge [] node {{\scriptsize $\begin{array}{l}
            \text{SearchLyric(song;artist;):SongRank;song;artist;ArtistUrl;}\\
            \text{SongUrl;lyricsId;lyricCheckSum}\\
            \ \\
            \text{SearchLyricText(lyricText;):SongRank;song;artist;ArtistUrl;}\\
            \text{SongUrl;lyricsId;lyricCheckSum}
            \end{array}$}}
            (Search)

            (Search) edge [loop right] node
                    {{\scriptsize $\begin{array}{l}
            \text{SearchLyric(song;artist;):SongRank;song;artist;}\\
            \text{ArtistUrl;SongUrl;lyricsId;lyricCheckSum}\\
            \ \\
            \text{SearchLyricText(lyricText;):SongRank;song;artist;}\\
            \text{ArtistUrl;SongUrl;lyricsId;lyricCheckSum}
            \end{array}$}} (Search)

            (Search) edge [] node
            {{\scriptsize $\begin{array}{l}
            \text{GetLyric(lyricCheckSum;lyricsId;):Lyric;LyricCorrectUrl;LyricRank;}\\
            \text{LyricCovertArtUrl;LyricCorrectUrl;artist;song}
            \end{array}$}} (end);
\end{tikzpicture}
\caption{LTS of the ChartLyrics service of Section \ref{sec:casestudy}.}
\label{fig:interfaceAutomaton}
\end{figure}

The formal model for reasoning about substitutability includes the behavioral protocols of both the expected and the actual services represented as Labelled Transition Systems (LTS) and formalized in PLTLB, in which each transition is labelled with the associated operation.
Input and output parameters of each operation are also part of the model (Fig. \ref{fig:interfaceAutomaton} shows the LTS of a service discussed in Section \ref{sec:casestudy}).

In addition, the model includes the definition of two kinds of integer counters.
The first is called {\em seen}, and it is used to check that the actual service can work using a subset of the input data provided by the client to the expected service.
The second is called {\em needed}, and it is used to check that the actual service
can provide a superset of the data the client expects to receive as output of
the expected service.
The model includes an instance of {\em seen} (resp. {\em needed}) for each type of data that can be used as input (resp. output) parameter for an operation.

The model states that each time an operation of the expected service is invoked, the instances of {\em seen} for each input parameter and those of {\em needed} for each output parameter are all \emph{incremented} by one.
%
Conversely, when an operation of the actual service is invoked, the instances of the {\em seen} counter for each input parameter and those of the {\em needed} counter for each output parameter are all \emph{decremented} by one.
Note that an actual service operation can be invoked only if the {\em seen} counter for each of its input parameters is $\geq 0$ (i.e. the input parameters have been provided by a client expecting to invoke some operations on the expected service).

Through this model, given a sequence of operations in the expected service automaton, we can formalize the problem of finding a substituting operation sequence in the actual service automaton.
More precisely, the actual operation sequence exists if, when the expected operation sequence is finished, the actual and expected services are in compatible states, and each instance of the {\em needed} counter has a value $\leq 0$.
The rationale behind the latter condition is that when the value of a {\em needed} counter is $0$, then the actual service provided enough instances of a certain type of data to fulfill client requests.
If, on the other hand, the actual service provides more instances of a type of data than those requested, then the corresponding {\em needed} counter is $< 0$.

\begin{figure}[htbp]
   \vspace{-.2cm}
    \begin{center}
   \includegraphics[width = \columnwidth]{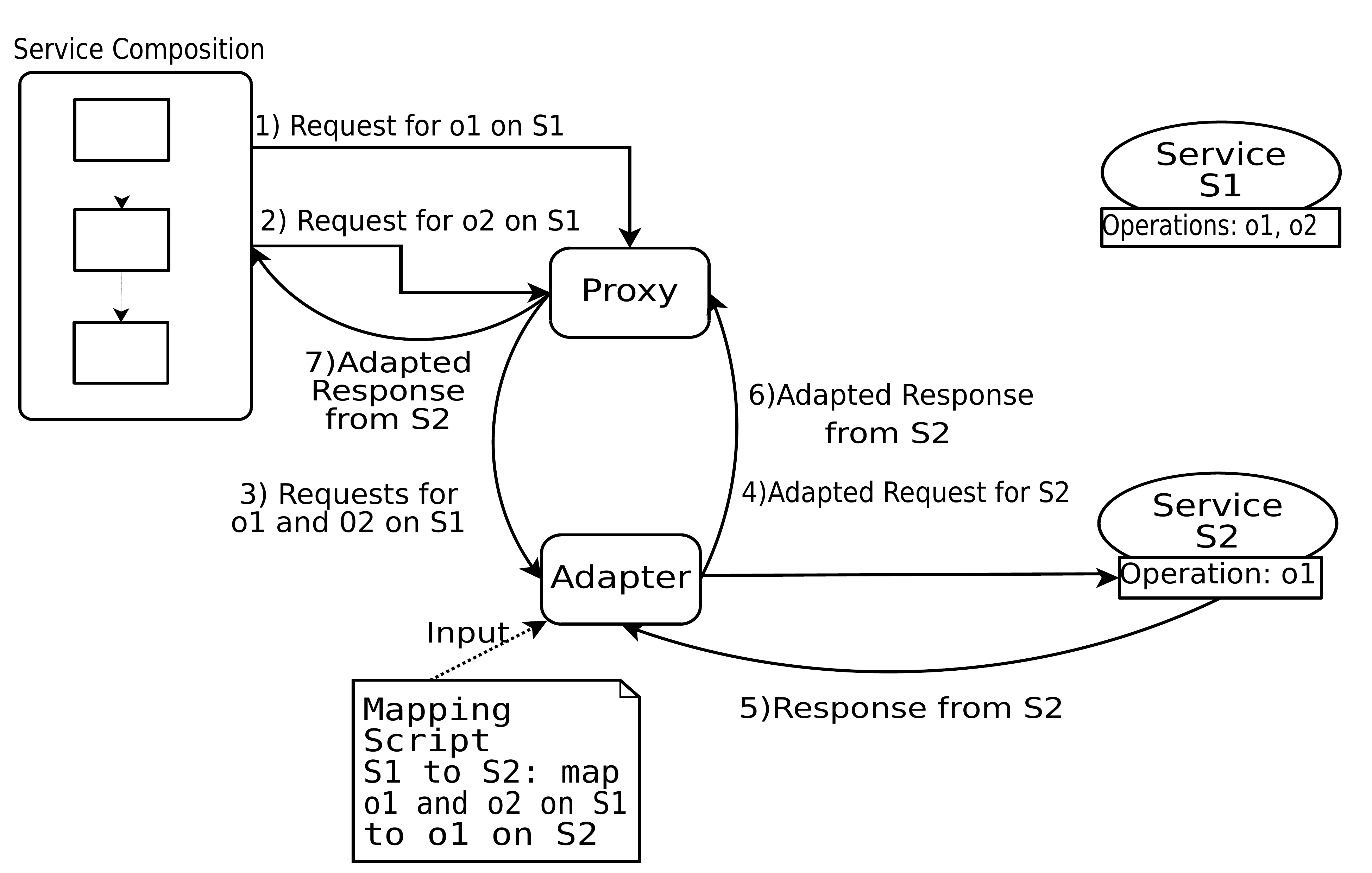}
    \caption{The adaptation runtime infrastructure.}
    \label{fig:proxy}
    \vspace{-.2cm}
    \end{center}
\end{figure}

In case the expected service operation sequence analyzed is substitutable by one
in the actual service, a mapping script is generated and then interpreted by an
\emph{adapter} that intercepts all service requests issued by the client and
transforms them into some requests the actual service can understand.
Fig. ~\ref{fig:proxy} shows the placement of adapters into the infrastructure
architecture and highlights their nature of intermediaries (see \cite{CNP09} for details).

\section{A logic for time-varying counters}
\label{sec:CLTL}

In order to deal with time-varying counters over
actual domains (such as \emph{seen} and \emph{needed} discussed in Section \ref{sec:applications}), we introduce an extension
of Linear-time Temporal Logic with past
operators and non-quantified first order integer variables.
The language we consider, denoted CLTLB(DL), is an
extension of PLTLB which combines pure
Boolean atoms and formulae with terms defined by DL constraints.
Counters can naturally be represented by integer variables over the whole domain
without any approximation due to a propositional encoding.
In \cite{BFMPRS10} we prove the decidability of the satisfiability problem in
more general cases.

Difference Logic is the structure $\langle \Z, =, (<_d)_{d \in \Z} \rangle$
where each $<_d$ is a binary relation defined as
\[
x <_d y \iFF x < y + d.
\]
The notations $x<y$, $x \leq y$, $x \geq y$, $x>y$ and $x=y+d$ are
abbreviations for $x <_0 y$, $x<_0y \vee x=y$, $\neg(x<_0 y)$,
$\neg(x<_0 y \vee x=y)$ and $y <_{d-1} x \wedge x <_{d+1} y$, respectively.

Let $AP$ the set of Atomic Propositions and $V$ the set of variables;
the CLTLB(DL) language is defined as follows:
\begin{equation*}
\begin{gathered}
  \phi :=
  \left\{
  \begin{gathered}
    p \mid \varphi \sim \varphi \mid \phi \wedge \phi \mid \neg \phi
\mid \\
   \X\phi \mid \Y\phi \mid \Zed\phi \mid \phi\,\U\phi \mid \phi\,\Snc\phi
  \end{gathered}
  \right.\\
  \varphi := x \mid \X \varphi \mid \Y \varphi
\end{gathered}
\end{equation*}
where $p \in AP$, $x \in V$, $\sim$ is any relation in DL, $\X$ is the usual
``next'', $\Y$, $\Zed$ are ``previous'' operators, $\U$ and $\Snc$
are the usual ``until'' and ``since'' operators.
Subformulae $\varphi$ are called \emph{arithmetic temporal terms} (a.t.t.);
for such terms, we define recursively the \emph{depth} $|\varphi|$: 
\[
\begin{gathered}
|x| = 0, \\ |X(\varphi)| = |\varphi| + 1, \\ |Y(\varphi)| = |\varphi| - 1.
\end{gathered}
\]
Depth extends naturally to formulae as the minimum depth of its a.t.t.'s.

The semantics of a formula $\phi$ of CLTLB(DL) is defined
w.r.t. a linear time structure $(S, s_0, I, \pi, L)$ where $S$ is the set of
states, $s_0$ is the initial state, $I:[|\phi|,-1]\times V\to \Z$ is an
assignment of variables, $\pi$ is an infinite path $\pi = s_0s_1\dots$
endowed with a sequence of valuations $\sigma: \N \times V \to \Z$
and $L: S \to 2^{AP}$ is the labeling function.
The function $I$ allows a valuation of variables to be defined also for instants
preceding zero and then to be extended to a.t.t.'s.
Indeed, if $\varphi$ is such a
term, $x$ is the variable in $\varphi$, $s_i$ is a state along the
sequence, and $\sigma^i$ is a shorthand for $\sigma(i,\cdot)$, then:
\begin{equation*}
\sigma^i(\varphi) = \left\{
                      \begin{array}{ll}
                        \sigma^{i+|\varphi|}(x), & \hbox{if $i+|\varphi|\geq
                         0$;} \\
                        I(i+|\varphi|,x), & \hbox{if $i+|\varphi|<0$.}
                      \end{array}
                    \right.
\end{equation*}
Given a model $\pi_\sigma$, the semantics of a formula $\phi$ is
recursively defined as:
\begin{equation*}
\begin{aligned}
\pi_\sigma^i \models p  &\iFF  p \in L(s_i) \text{ for } p \in AP \\
\pi_\sigma^i \models (\varphi_1 \sim \varphi_2) &\iFF
  \sigma^{i+|\varphi_1|}(x_{\varphi_1}) \sim \sigma^{i+|\varphi_2|}(x_{\varphi_2}) \\
\pi_\sigma^i \models \neg \phi &\iFF  \pi_\sigma^{i} \not\models \phi \\
\pi_\sigma^i \models \phi \wedge \psi &\iFF  \pi_\sigma^i \models \phi
\, \text{and} \, \pi_\sigma^i \models \psi\\
\pi_\sigma^i \models \X \phi &\iFF \pi_\sigma^{i+1} \models \phi \\
\pi_\sigma^i \models \Y \phi &\iFF \pi_\sigma^{i-1} \models
\phi \wedge i>0\\
\pi_\sigma^i \models \Zed \phi &\iFF \pi_\sigma^{i-1}
\models \phi \vee i=0\\
\pi_\sigma^i \models \phi\U\psi &\iFF
\left\{ \begin{gathered}
  \exists \, j\geq i: \pi_\sigma^j \models \psi \, \wedge \\ \pi_\sigma^n
\models \phi \, \forall\, i\leq n < j
\end{gathered}\right. \\
\pi_\sigma^i \models \phi\Snc\psi &\iFF
\left\{\begin{gathered}
\exists \, 0\leq j \leq i: \pi_\sigma^j \models \psi \, \wedge \\
\pi_\sigma^n \models \phi \, \forall\, j < n \leq i
\end{gathered}\right. \\
\end{aligned}
\end{equation*}
where $x_{\varphi_i}$ is the variable that appears in $\varphi_i$ and
$\sim$ is any relation in DL.
The $\R$ and $\T$ operators, over infinite paths, can be defined as
usual: $\phi\R \psi \equiv \neg(\neg\phi\U \neg\psi)$ and $\phi\T \psi \equiv \neg(\neg\phi\Snc \neg\psi)$.
By means of previous dualities and DeMorgan's rules, it is always possible to rewrite all
formulae to \emph{positive normal form}.
From now on, we assume all formulae are in positive normal form.
A formula $\phi \in$ CLTLB(DL) is \emph{satisfiable} if there exists a linear
time structure $(S, s_0, I, \pi, L)$ and a sequence of valuations $\sigma$ such that $\pi_\sigma^0 \models
\phi$; where $\pi_\sigma^0$ is the the sequence built from $\pi$
and the valuations as described before.

Unfortunately, CLTLB(DL) is too expressive in the sense that the
satisfiability problem can be proven to be highly undecidable
\cite{CC00}.
However, the satisfiability and the
model checking problems for a CLTLB(DL) formula $\phi$ for
$k$-\emph{partial} valuations (i.e., for all computation
in which the value of counters is considered only up to $k$ plus the \emph{maximum} depth of the subformulae of $\phi$ steps) is shown to be decidable \cite{BFMPRS10}.
Both of them reduce to the satisfiability and the model checking problems, respectively,
over bounded paths of length equal to $k$ with $k$-partial valuations.
As in the standard BMC (of a property $\phi$)
the goal is looking for finite initialized path of the system that are
witnesses of wrong behaviors, i.e., paths along which the negations of
the property $\phi$ holds.
When the finite path of length $k$ admits
a loop, it contains all its infinite periodic behavior; and
conversely, when a loop does not exists, it represents all its possible
extensions.
Indeed, it is representative of an infinite path.
Formally, paths are words of states $s_i$ which may be possibly
periodic: $\pi=uv^\omega$ with $u=s_0\cdots s_l$ and
$v=s_{l+1}\cdots s_k$ where $l\leq k$, if the loop exists;
$\pi=uv$, if it does not.
Beside the propositional model, the \emph{values} of the
variables up to the state $s_k$ are depicted by a bounded
representation $\pi_{\sigma_k}$ of the model $\pi_{\sigma}$.
It is also opportunely bordered by some values
of variables referring to time instants outwards the finite path, before $s_0$
and after $s_k$ depending on the depth of the formula.
Arithmetic DL constraints may be part of the possibly periodic model
$\pi_\sigma$ and, thus, are defined by means of a finite prefix of length $k$.
According to \cite{BFMPRS10,BCCSZ03}, we are allowed
to use a proper bounded semantics to state
reachability properties on that part of the system involving a counting
mechanism (i.e., $\X x = y + 1$, where $x$, $y$ are variables).
Note that over finite acyclic paths, the equivalence  $\phi \R \psi
\equiv \neg(\neg\phi \U \neg\psi)$ and $\phi \T \psi \equiv
\neg(\neg\phi \Snc \neg\psi)$ no longer holds.
Then, $\R$ (and symmetrically $\T$) is redefined as \cite{BHJLS06}:
\[
\begin{gathered}
\pi_\sigma^i \models_k \phi \R \psi \iFF\\
\exists \,  i\leq j \leq k, \pi_\sigma^j \models_k \phi \, \wedge
\pi_\sigma^n \models_k \psi \, \forall\, i \leq n \leq j
\end{gathered}
\]

Based on this assumption, the (existential) reachability problem
over infinite path endowed with a $k-$partial valuation $\sigma_k$, $\pi_{\sigma_k} \models \phi$, can be
reduced to the bounded (existential) reachability problem over finite
paths (possibly cyclic) with $k-$partial valuation $\pi_{\sigma_k} \models_k \phi$:
\begin{theorem}[\cite{BFMPRS10}]
Let $\phi$ be a \emph{CLTLB(DL)} formula. There exists $k>0$ such
that if $\pi_{\sigma_k}$ is a path endowed
with a $k$-partial valuation of variables, then $\pi_{\sigma_k} \models \phi
\iFF \pi_{\sigma_k} \models_k \phi$.
\end{theorem}
These results allow us to correctly verify the satisfiability of CLTLB(DL)
formulae and also to realize a bounded model checking of systems involving DL
constraints.
Particularly, when a counting mechanism is defined, reachability properties of
values of variables along paths of finite length can be verified.
Obviously, if the reachability property does not hold within $k$, then $k$ can
be refined and augmented.
As explained later in Section \ref{sec:experimental}, the substitutability
problem can be significantly solved by means of a BMC approach by correctly
estimating an upper bound of $k$.
This is done by using an opportune heuristic based
on the dimension of the automata describing services and the length of traces of
invocations.
For this reasons, the substitutability problem, which requires to check counting
mechanism over finite paths of invocations of service functions, can be easily
encoded to a bounded reachability problem. 

\section{Encoding of Bounded Reachability Problem}
\label{sec:encoding}

In this section the bounded reachability problem is encoded as the
satisfiability of a Quantifier Free Integer Difference Logic formula
with Uninterpreted Function and predicate symbols (QF-UFIDL).
Such a logic is shown to be decidable,
and the satisfiability problem to be {\sc NP}-complete, as it can be easily proved applying Nelson-Oppen Theorem.
The QF-UFIDL encoding results to be more succinct and expressive than the
Boolean one: lengthy propositional constraints are substituted
by more concise DL constraints and arithmetic (infinite) domains do
not require an explicit finite representation.
These facts, considering also that the satisfiability problem for QF-UFIDL has the same complexity of SAT, make the SMT-based approach
particularly efficient to solve runtime substitutability problem, as
demonstrated by performance results.
In the key work of Biere et al. \cite{BCCSZ03}, the BMC is reduced to a pure
propositional satisfiability problem.
This approach, and further refinements \cite{BHJLS06,PMS07,PMS09}, has been already
implemented in the Zot tool\footnoteremember{Zot}{Zot: a Bounded Satisfiability Checker, http://home.dei.polimi.it/pradella/}.

\subsection{Encoding the Time}
As discussed before, the BMC problem amounts to look for a finite
representation of infinite (possibly periodic) paths.
The SAT-based approach encodes finite paths \cite{BCCSZ03} by means of $2k+3$
propositional variables. The time instant at which the periodic suffix
starts is defined by the {\em loop selector
variables} $l_0, l_1, \dots l_k$: $l_i$ holds if and only if the loop
starts at instant $i$, i.e., $s_{i}$ is the successor of $s_k$.
Then, the truth (of atomic proposition) in $s_i$ and $s_k$,
defined by the labeling function $L$ defined in Section \ref{sec:CLTL}, must be the same.
Further propositional variables, $\mathit{inLoop}_i$ ($0 \le i\le k
$) and $\mathit{loopEx}$, respectively, mean that time instant $i$ is
inside a loop and that there actually exists a loop.

The same temporal behavior can be defined by means of \emph{one} QF-UFIDL
formula involving only \emph{one} integer
\emph{loop-selecting} variable $\bm{loop} \in \mathbb{Z}$:
\[
\bigwedge_{i=1}^k \left( \bm{loop}=i \Rightarrow L(s_{i-1}) = L(s_k) \right ).
\]
The QF-UFIDL encoding is more concise: it does not require $2k+3$
Boolean variables ($l_i$, $inLoop_i$ and $loopExists$). A value of
$\bm{loop}$ between $1$ and $k$ defines if there exists a loop and its
position; it does not depend on the $k$ parameter.

\subsection{Encoding the Arithmetic Temporal Terms}
Since CLTLB(DL) formulae consist also of a.t.t.'s, we need to
define a suitable semantics for them.
An \textit{arithmetic formula function}, i.e. an uninterpreted function
$\bm{\alpha}: \Z \to \Z$, is associated with each arithmetic temporal subterm of
$\Phi$.
Let $\alpha$ be such a subterm, then the arithmetic
formula function associated with it (denoted by the same name but in written
in bold face), is recursively
defined w.r.t. the sequence of valuations $\sigma$ as:
\[
\begin{array}{c|c}
    \alpha & 0 \leq i \leq k
    \\
    \hline
    x & \bm{x}(i) = \sigma^i(x) \\
    \X\alpha & \,\,\bm{\X\alpha}(i) = \bm{\alpha}(i+1)    \\
    \Y\alpha & \,\,\bm{\Y\alpha}(i) = \bm{\alpha}(i-1)   \\
\end{array}
\]
This semantics is well-defined between $0$ and $k$ thanks to the
initialization function $I$.

\subsection{Encoding the Propositional Terms}
The propositional encoding is inspired from that one studied in
\cite{BHJLS06} but deeply revised to take also into account relations over
a.t.t.'s.
In the case of Boolean encoding, the semantics of a PLTLB formula $\Phi$ is
defined w.r.t. the truth value of all its
subformulae only by means of
Boolean variables $t$ associated to each of them,
for all $0\leq i \leq k+1$: if $t_i$ holds then the subformula
$t$ holds at instant $i$.
The instant $k+1$ is appended to the path to easily represent the instant in the past
where the loop realizes the periodicity; indeed, it turns to be useful for the encoding.
The propositional semantics of a CLTLB(DL) formula
$\Phi$ is defined alike that one of PLTLB.
The QF-UFIDL encoding, instead, associates to each propositional subformula a
\textit{formula predicate} that is a unary uninterpreted predicate
$\bm{\varphi} \in \mathcal{P}(\Z)$.
When the subformula $\varphi$ holds at instant $i$ then $\bm{\varphi}(i)$ holds.
As the length of paths is fixed to $k+1$, and all paths start from $0$,
formula predicates are actually subsets of $\{0, \dots, k+1\}$.
Let $\varphi$ be a propositional subformula of $\Phi$, $\alpha$,
$\beta$ be two a.t.t.'s and $\sim$ be any relation in DL; then the formula predicate
associated with $\varphi$ (denoted by the same name but written in bold face), is recursively defined as:
\[
\begin{array}{c|c}
  \varphi & 0 \leq i \leq k+1 \\
  \hline
  p & \,\,\qquad\bm{p}(i) \iFF p \in L(s_i) \\
  \alpha \sim \beta & \,\,(\bm{\alpha \sim \beta})(i) \iFF \bm{\alpha}(i) \sim
\bm{\beta}(i)\\
  \neg \phi &  \bm{\neg\phi}(i) \iFF \neg \bm{\phi}(i)\\
  \phi \wedge \psi & \,\,(\bm{\phi \wedge \psi})(i) \iFF \bm{\phi}(i)
\wedge \bm{\psi}(i)
\end{array}
\]

\subsection{Encoding Temporal Operators}
{\em Temporal subformulae constraints} define the basic temporal
behavior of future and past operators, by using their traditional
fixpoint characterizations. Let $\phi$ and $\psi$ be
propositional subformulae of $\Phi$, then:
\[
\begin{array}{c|c}
  \varphi & 0 \leq i \leq k \\
  \hline
  \X\phi & \bm{\X\phi}(i) \iFF \bm{\phi}(i+1) \\
  \phi\U\psi & (\bm{\phi\U\psi})(i)\iFF(\bm{\psi}(i) \vee (\bm{\phi}(i) \wedge
(\bm{\phi\U\psi})(i+1)))\\
  \phi\R\psi & (\bm{\phi\R\psi})(i)\iFF(\bm{\psi}(i) \wedge (\bm{\phi}(i) \vee
(\bm{\phi\R\psi})(i+1)))\\
\end{array}
\]
\[
\begin{array}{c|c|c}
  \varphi & 0 < i \leq k+1 & i = 0\\
  \hline
  \Y\phi & \bm{\Y\phi}(i) \iFF \bm{\phi}(i-1) & \neg \bm{\Y\phi}(0)\\
  \Zed\phi & \bm{\Zed\phi}(i) \iFF \bm{\phi}(i-1) & \bm{\Zed\phi}(0)\\
  \phi\Snc\psi &
  \begin{array}{c}
    (\bm{\phi\Snc\psi})(i)\iFF(\bm{\psi}(i)\vee \\ (\bm{\phi}(i) \wedge
(\bm{\phi\Snc\psi})(i-1)))
  \end{array}&
  \begin{array}{c}
  (\bm{\phi\Snc\psi})(0) \iFF\\ \bm{\psi}(0)
  \end{array}\\
  \phi\T \psi &
  \begin{array}{c}
    (\bm{\phi\T \psi})(i)\iFF(\bm{\psi}(i) \wedge \\ (\bm{\phi}(i) \vee
(\bm{\phi\T\psi})(i-1)))
  \end{array}&
  \begin{array}{c}
    (\bm{\phi\T\psi})(0) \iFF \\ \bm{\psi}(0)
  \end{array}
\end{array}
\]
{\em Last state constraints} define an equivalence between
truth in $k+1$ and those one indicated by $\bm{loop}$, since the instant
$k+1$ is representative of the instant $\bm{loop}$ along periodic paths.
Otherwise, truth values in $k+1$ are trivially false.
These constraints have a similar structure to the
corresponding Boolean ones, but here
they are defined by only \emph{one} DL constraint, for each
subformula $\varphi$ of $\Phi$,  w.r.t. the variable $\bm{loop}$:
\[
  \begin{array}{l}
    \left(\bigwedge_{i=1}^k (\bm{loop}=i \Rightarrow (\bm{\varphi}(k+1) \iFF \bm{\varphi}(i)))\right) \wedge \\
    \left(\left(\bigwedge_{i=1}^k \neg(\bm{loop}=i)\right) \Rightarrow (\neg
\bm{\varphi}(k+1))\right).
  \end{array}
\]
Note that if a loop does not exists then the fixpoint semantics of
$\R$ is exactly that one defined over finite acyclic path in Sec. \ref{sec:CLTL}.
Finally, to correctly define the semantic of $\U$ and $\R$, their
\emph{eventuality} have to be accounted for.
Briefly, if $\phi\U\psi$ holds at $i$, then $\psi$ eventually holds
in $j\geq i$; if $\phi\R\psi$ does not hold at $i$, then $\psi$
eventually does not hold in $j\geq i$.
Along finite paths of length $k$, eventualities must hold between $0$
and $k$.
If a loop exists, an eventuality may holds within the loop.
The original Boolean encoding introduces $k$ propositional variables
for each $\phi\U\psi$ and $\phi\R\psi$ subformula
of $\Phi$, for all $1 \leq i \leq k$, which represent the eventuality of $\psi$
implicit in the formula.
The interested reader should consult \cite{BHJLS06}.
Differently, in the QF-UFIDL encoding, only \emph{one} variable $\bm{j_\psi} \in
\mathbb{Z}$
is introduced for each $\psi$ occurring in a subformula
$\phi\U\psi$ or $\phi\R\psi$.
\[
\begin{array}{c|c}
    \varphi & \mathrm{Base} \\
    \hline
    \phi\U\psi &
    \begin{array}{l}
    \left(\bigvee_{i=1}^k \bm{loop}=i\right) \Rightarrow\\
    \qquad \qquad (\bm{\varphi}(k) \Rightarrow
     \bm{loop} \leq \bm{j_\psi} \leq k \wedge \bm{\psi}(\bm{j_\psi}))
    \end{array}
    \\
    \phi\R\psi &
    \begin{array}{l}
    \left(\bigvee_{i=1}^k \bm{loop}=i\right) \Rightarrow\\
     \qquad (\neg\bm{\varphi}(k) \Rightarrow
     \bm{loop} \leq \bm{j_\psi} \leq k \wedge \neg \bm{\psi}(\bm{j_\psi}))
    \end{array}
  \end{array}
\]
The complete encoding of $\Phi$ consists of the logical
conjunction of all above components, together with $\Phi$ evaluated at the
first instant along the time structure.

Let $\Phi$ be a pure propositional formula, actually in PLTLB, then we can
compare the dimension of the SAT-based encoding versus the QF-UFIDL one.
If $m$ is the total number of subformulae and $n$ is the total number of
temporal operators $\U$ and $\R$ occurring in $\Phi$, then the SAT-based
encoding requires $(2k+3) + (k+2)m + (k+1)n = O(k(m+n))$ fresh propositional
variables.
Differently, the QF-UFIDL encoding
requires only $n+1$ integer variables ($\bm{loop}$ and $\bm{j_\psi}$) and $m$ unary
predicates (one for each subformula).

\section{Case Study}
\label{sec:casestudy}
To demonstrate our methodology, we use an example concerning two existing conversational services available on the Internet. These two services realize two lyric search engines. One is
called {\em ChartLyrics}~\footnote{http://www.chartlyrics.com/api.aspx}, the other {\em LyricWiki}~\footnote{http://lyrics.wifkia.com/Main\_Page}.

{\em ChartLyrics} is a lyrics database sorted by artists or songs. The WSDL~\footnote{http://api.chartlyrics.com/apiv1.asmx?WSDL} of {\em ChartLyrics} provides three operations: (i) {\em SearchLyric} to search available lyrics, (ii) {\em SearchLyricText} to search a song by means of some text within an available lyric text, and (iii) {\em GetLyric} to retrieve the searched lyric.

\begin{figure}\centering
\begin{tikzpicture}[node distance=4cm,auto,
                    every state/.style={draw=black!100,scale=0.5}, >=stealth]

  \node[state] (s4) {{\huge $s_4$}};
  \node[state] (s1) [above of=s4] {{\huge $s_1$}};
  \node[state] (s2) [right of=s1] {{\huge $s_2$}};
  \node[state] (s3) [right of=s2] {{\huge $s_3$}};
  \node[state] (s5) [right of=s4] {{\huge $s_5$}};
  \node[state] (s6) [right of=s5] {{\huge $s_6$}};
  \node[state] (start) [left of=s4, yshift=2cm] {{\huge start}};

  \path[->] (start) edge [] node {{\tiny \text{cSE(1;0):2;4}}} (s1)
            (start) edge [] node[below left] {{\tiny
\text{sS(0;1):1;0}}} (s4)
            (s1) edge [] node {{\tiny \text{sA(10):0}}} (s2)
            (s1) edge [] node[below left] {{\tiny \text{sS(0;1):1;0}}}
(s5)
            (s4) edge [] node {{\tiny \text{cSE(1;0):2;4}}} (s5)
            (s2) edge [] node[above] {{\tiny \text{gA(0):0;3;5;6;7}}} (s3)
            (s5) edge [] node[above] {{\tiny $\begin{array}{l}
                                                           \text{sA(10):0} \\
                                                           \text{gS(2;4;0;1):8;9;1;0}
                                                         \end{array}$}} (s6)
            (s3) edge [] node[right] {{\tiny \text{sS(0;1):1;0}}} (s6)
            (s2) edge [] node[below left] {{\tiny \text{sS(0;1):1;0}}} (s6)

            (s1) edge [loop above] node {{\tiny \text{cSE(1;0):2;4}}} (s1)
            (s2) edge [loop above] node {{\tiny $\begin{array}{l}
                                                           \text{sA(10):0} \\
                                                           \text{cSE(1;0):2;4}
                                                         \end{array}$}} (s2)
            (s3) edge [loop right] node {{\tiny $\begin{array}{l}
                                                           \text{sA(10):0} \\
                                                           \text{cSE(1;0):2;4}
                                                         \end{array}$}} (s3)
            (s4) edge [loop below] node {{\tiny \text{sS(1;0):2;4}}} (s4)
            (s5) edge [loop below] node {{\tiny $\begin{array}{l}
                                                           \text{sS(0;1):1;0} \\
                                                           \text{cSE(1;0):2;4}
                                                         \end{array}$}} (s5)
            (s6) edge [loop right] node {{\tiny $\begin{array}{l}
                                                           \text{sA(10):0} \\
                                                           \text{sS(0;1):1;0} \\
                                                           \text{cSE(1;0):2;4} \\
                                                           \text{gS(2;4;1):8;9;1;0} \\
                                                           \text{gA(0):0;3;5;6;7}
                                                         \end{array}$}} (s6);

\end{tikzpicture}
\caption{A subset of behavior protocol automaton of {\em
LyricWiki}. \textbf{Operations}: searchSongs (sS), checkSongExists (cSE),
searchArtists (sA), getArtist (gA), getSong (gS). \textbf{Parameters}:
artist (0), song (1), lyricsId (2), item (3), lyricCheckSum (4), SongUrl (5), year (6),
album (7), LyricCorrectUrl (8), Lyrics (9), lyricText (10). }
\label{fig:interfaceAutomaton2}
\end{figure}
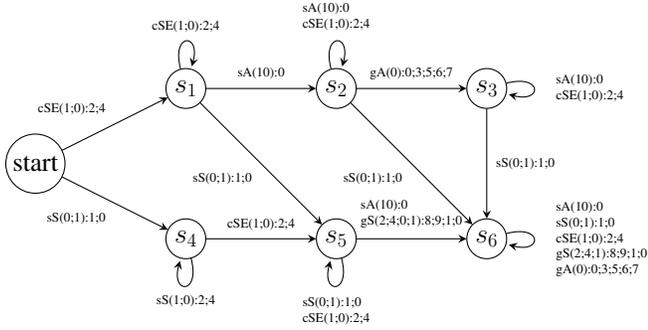
{\em LyricWiki} is a free site where anyone can go to get reliable lyrics for any song from any artist. The WSDL of {\em LyricWiki}~\footnote{http://lyrics.wikia.com/server.php?wsdl} provides several operations.
Five of them are of interest for our purposes: (i) {\em searchSongs} to search for a possible song on {\em LyricWiki} and get up to ten close matches, (ii) {\em checkSongExists} to check if a song exists in the {\em LyricWiki} database, (iii) {\em getSong} to get the
lyrics for a searched {\em LyricWiki} song with the exact artist and song match, (iv) {\em searchArtists} to search for a possible artist
by name and return up to ten close matches, and (v) {\em getArtist} to get the entire discography for a searched artist.
To get a lyric through {\em ChartLyrics}, a client can exploit the following sequence of operation invocations: {\em
SearchLyric}, {\em GetLyric}.
Conversely, to get a lyric through {\em LyricWiki}, a possible sequence of operation invocations is the
following: {\em checkSongExists}, {\em searchSongs}, {\em getSong}
(see the representation of the conversational protocols of {\em ChartLyrics} and {\em LyricWiki}, respectively, in
Fig. \ref{fig:interfaceAutomaton} and Fig. \ref{fig:interfaceAutomaton2}).

If {\em LyricWiki} were part of a web application realized through a service composition, it could happen that, in certain circumstances, it would need to be replaced by {\em ChartLyrics} or by any other specialized search engine.
This could happen, for instance, to accommodate the preferences of users having their preferred engine, or to handle the cases when {\em LyricWiki} is unavailable for any reason.
The developer could code, by hand, the instructions to deal with any possible engine and its replacement.
However, this approach does not allow the application to deal with search engines unknown at design time.
A better solution, which would overcome this problem, is to build a mapping mechanism that dynamically handles the mismatches by automatically synthesizing a behavior protocol mapping script. The adaptation realized by the synthesized mapping script could state, e.g., that the sequence of {\em LyricWiki} operations {\em checkSongExists}, {\em searchSongs}, {\em getSong} maps to the sequence of {\em ChartLyrics} operations {\em SearchLyric}, {\em GetLyric}. 

Let us consider as an example the expected service operation sequence
{\em checkSongExists}, {\em searchSongs}, {\em getSong}, which brings
the {\em LyricsWiki} behavior protocol automaton from state $ start$
to state $ s_6$ (see Fig. ~\ref{fig:interfaceAutomaton2}).
We assume to have established a compatibility relation
between services' data. Also, for the sake of brevity, the automata of Fig. \ref{fig:interfaceAutomaton} and~\ref{fig:interfaceAutomaton2} are represented with this relation already established, though in practice this requires an additional mapping step (for more details see \cite{cavallaro-pesos-09, seams10Adapt}).
Finally, we establish a state compatibility relation.
This defines that state $s_6$ of the expected service is compatible with state {\em end} of the actual service, which means that if the  expected service reaches state $s_6$, then the actual service should reach state $end$. The example expected operations sequence starts from the $start$ state and leads the behavior protocol model into state $s_6$. 

The automata describing service protocols, the state compatibility relation and the expeced service operation sequence are all formalized through suitable CLTLB(DL) formulae $expectedService$, $actualService$ and $expectedOperationSequence$.
Then, we formulate the problem of checking if the expected service can be substituted by the actual service in terms of a bounded reachability problem over the automata describing the protocols of the expected and actual services. 
The problem consists in searching for a finite operation sequence on the actual service automaton which starts (resp. ends) in a state compatible with the start (resp. end) state of the expected service operation sequence.
Moreover, the actual service operation sequence should require no more input parameters than those provided to the expected service sequence, and it should provide at least the same parameters provided by the expected service sequence.
To ensure this property we keep track, through instances of counters \emph{seen} and \emph{needed} (see Section \ref{sec:applications}), of how many parameters of any given kind are provided as input to the expected service operations and of how many parameters of any given kind are returned by each actual service operation (this is formalized through suitable CLTLB(DL) formulae $seen$ ad $needed$).
Finally, a solution for the bounded reachability problem can be obtained by checking the satisfiability of CLTLB(DL) formula $expectedService \wedge actualService \wedge expectedOperationSequence \wedge seen \wedge needed$.

Considering the example sequence on  {\em LyricsWiki}, a client expecting to invoke this sequence is assuming to provide as input to the first operation of the sequence a song and an artist. This will set the {\em seen} counter to $1$ for both provided inputs.
Moreover, it expects the invoked operation to return a {\em lyricsId} and a {\em lyricCheckSum}, which will increment the corresponding instances of the {\em needed} counter to $1$. Considering the actual service protocol, our approach searches for an operation accepting a subset of  the provided input data and providing a superset of the required return data.

The operation to be selected should leave the $start$ state as the state compatibility relation provided as input for the approach mandates the compatibility of state \emph{start} of {\em LyricsWiki} with state \emph{start} of {\em ChartLyrics}.
In our example the invocation of {\em checkSongExists} makes {\em SearchLyric} the only suitable candidate.
After the invocation of this actual service operation all instances of {\em seen} and those instances of {\em needed} associated to theoutput parameters of {\em checkSongExists} are reset to $0$.
The actual service operation returns also some extra data that are not required by the invoked expected service operation (i.e. song, artist, songRank, artistUrl, songUrl).
In this case the reasoning mechanism offers two possible choices: extra data can be discarded (hence ignored also in the future), or they can be initially ignored, but stored for an eventual later use.
The former strategy is more conservative, but it may also limit the possibility of the reasoning mechanism to find an adapter.
The latter strategy may affect data consistency in some cases, as it allows using as a reply for an operation some data that have been received before the request has been actually issued, but it also opens the possibility of finding adapters in situations in which the former would fail.
In this case study we use the latter strategy, hence the {\em needed} counters for those data that are not required as a response by the invoked expected service operation are set to $-1$.
\begin{table*}[t!]
 \centering
\scriptsize
\begin{tabular}{|c|l|l|}
\hline
\emph{Step}   &  \emph{ Execution trace Content } & Counters value \tabularnewline
\hline
 \multirow{3}{*} {1} &  {\em LyricWiki}State:start ; {\em LyricWiki}Operation:checkSongExists & \multirow{3}{*} {All counters set to 0} \tabularnewline
 & {\em LyricWiki}Input: song, artist; {\em LyricWiki}Output:lyricId, lyricCheckSum   &\tabularnewline
 &{\em chartLyrics}State:start; {\em LyricWiki}Operation:checkSongExists  & \tabularnewline
\hline
\multirow{4}{*} {2}  & {\em LyricWiki}State:$s_1$ & \multirow{2}{*} {seen(song) = seen(artist) = 1}\\
&  {\em chartLyrics}Input: song, artist & \multirow{4}{*} {needed(lyricId) = needed(lyricCheckSum) = 1}\\
& {\em chartLyrics}Output:song , artist, artistUrl, songRank, lyricsId, lyricChecksum &\\
& {\em chartLyrics}State:start; {\em chartLyrics}Operation:searchLyric &\\
\hline
\multirow{4}{*} {3} & {\em LyricWiki}State:$s_1$; {\em LyricWiki}Operation:searchSongs &seen(song) = seen(artist) = 0\\
&  {\em LyricWiki}Input:song, artist;  {\em LyricWiki}Output:song, artist &needed(lyricsId) = needed(lyricCheckSum) = 0\\
& {\em chartLyrics}State:searchLyric\_start & needed(artist) = needed(artistUrl) =  -1\\
& &needed(song) = needed(songRank)= -1\\
\hline
\multirow{3}{*} {4}   & {\em LyricWiki}State:$s_5$ & seen(song) = seen(artist) = 1\\
  & {\em chartLyrics}State:searchLyric\_start  &needed(song) = needed(artist) = 0\\
  & {\em chartLyrics}Operation: {\em None} &\\
\hline
\multirow{4}{*} {5}   &{\em LyricWiki}State:$s_5$; {\em LyricWiki}Operation: getSong  &\multirow{4}{*} {No Changes}\\
&{\em LyricWiki}Input: lyricId, song, lyricCheckSum, artist &\\
  &{\em LyricWiki}Output:song, artist, lyricCorrectUrl, Lyric &\\
  &{\em chartLyrics}State:searchLyric\_start &\\
\hline
\multirow{5}{*} {6}  &   {\em LyricWiki}State:$s_6$&seen(song) = seen(artist) = 2\\
& {\em chartLyrics}Input: lyricId, lyricCheckSum &  seen(lyricCheckSum) = seen(lyricId) = 1\\
  & {\em chartLyrics}Output: song , artist, artistUrl, lyricRank, Lyric, lyricCorrectUrl, lyricCoverArtUrl &needed(song) = needed(artist) = 1\\
  & {\em chartLyrics}State:searchLyric\_start {\em chartLyrics}Operation:getLyric &needed(lyricCorrectUrl) = needed(Lyric) = 1\\
  \hline
\multirow{4}{*} {7} & {\em LyricWiki}State:$s_6$ &seen(lyricCheckSum) = seen(lyricId) = 0 \\
  &{\em LyricWiki}Operation: {\em None}  &needed(song) = needed(artist) = 0\\
  & {\em chartLyrics}State:end&needed(lyricCorrectUrl) = needed(Lyric) = 0 \\
  & {\em chartLyrics}Operation: {\em None}& needed(artistUrl) = needed(lyricRank) = -1 \\
\hline

\end{tabular}
\normalsize
\caption{Mapping script generated for the example in this section} \label{tab:sequence} \vspace{-0.5cm}
\end{table*}

After the invocation of {\em SearchLyric} the actual service goes in {\em SearchLyric\_start} state.
The next operation on the expected sequence to be invoked is {\em searchSongs}, which requires as input the names of the song to be searched and of its author and provides as return parameters the names of the artist and of the song, if they are found.
Since the {\em needed} counters for both the name of the artist and of the song are set to $-1$, instances of those data have been previously stored, hence no operation shall be invoked on the actual service, which remains in state {\em SearchLyric\_start}.

The last operation in the expected sequence is {\em getSong}, which requires as input artist and song names and the id and checkSum returned by the previously invoked {\em checkSongExists}. The expected service has again the same three operations of the previous step available, but this time there are two available candidates for selection: {\em searchSongs} and {\em GetLyric}. In this situation the latter is selected, because of the state compatibility relation provided as input to the adapter search phase.  Given the data-flow constraints elicited before,  {\em GetLyric} is the only available operation that can satisfy also the state compatibility relation. After the invocation of {\em GetLyric} the expected and actual services are in compatible states and the {\em needed} counter instances are all set to $0$.
Then, the actual service operation sequence found can be substituted to the expected service sequence.

A mapping script generated  for the example sequence in this section is reported in Table~\ref{tab:sequence}.
Each step contains the state in which each one of the analyzed automata is, the operations in $seq_{{exp}}$ and in $seq_{{act}}$ that should be invoked in that step, and the exchanged data, if any.
For each operation in $seq_{{exp}}$ the adapter expects to receive an invocation for the expected service, and for each operation in $seq_{{act}}$ the adapter performs an invocation to the actual service.
The table shows also the updates for the {\em seen} and {\em needed} counters.

\section{Evaluation and Experimental Results}
\label{sec:experimental}

\begin{figure*}[t!]
 \centering
 \subfigure[Elapsed times on the second set of experiments]{
   \includegraphics[width=.9\columnwidth]{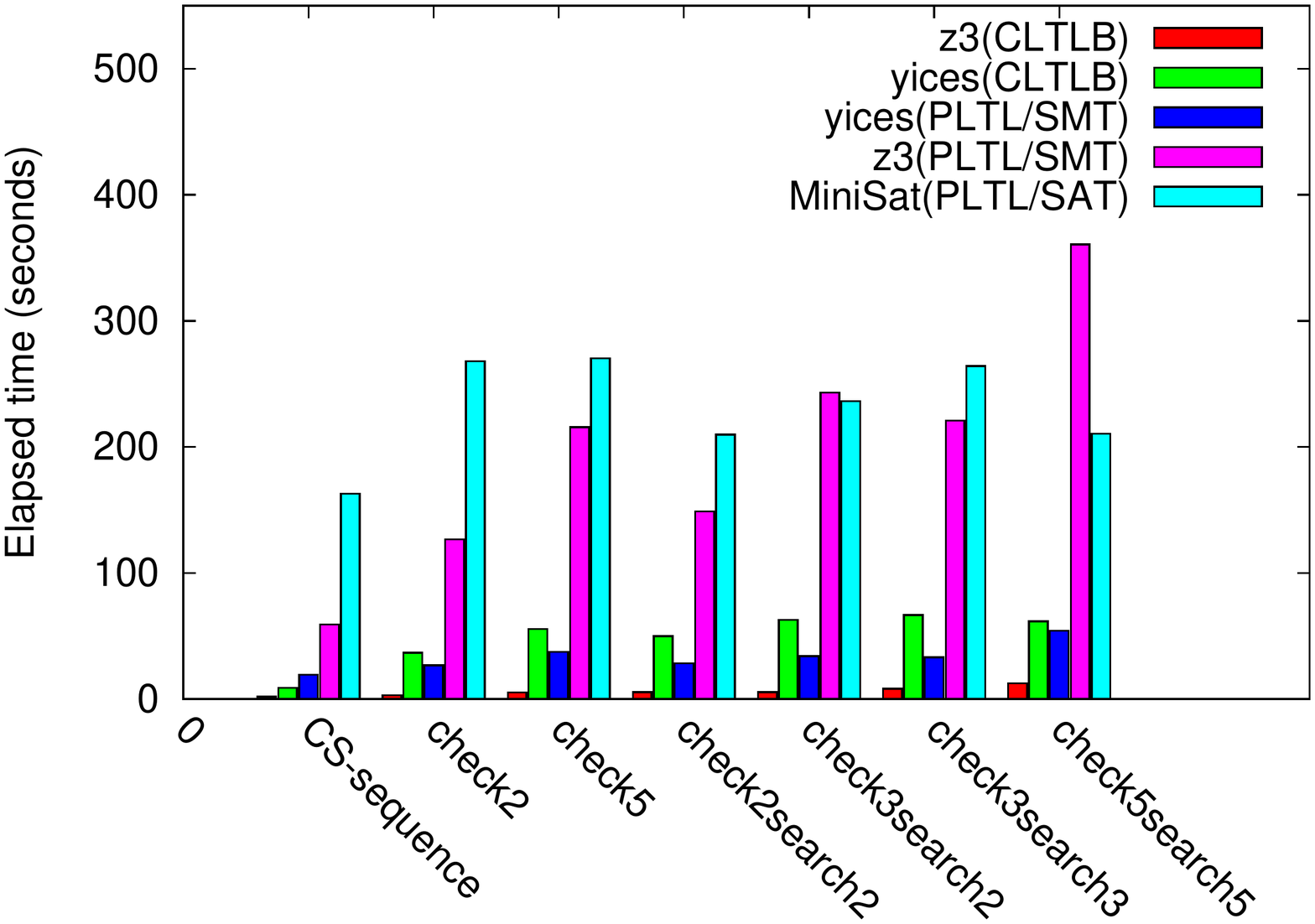}
   \label{fig:realcaseT}
 }
 \subfigure[Memory occupations on the second set of experiments]{
   \includegraphics[width=.9\columnwidth]{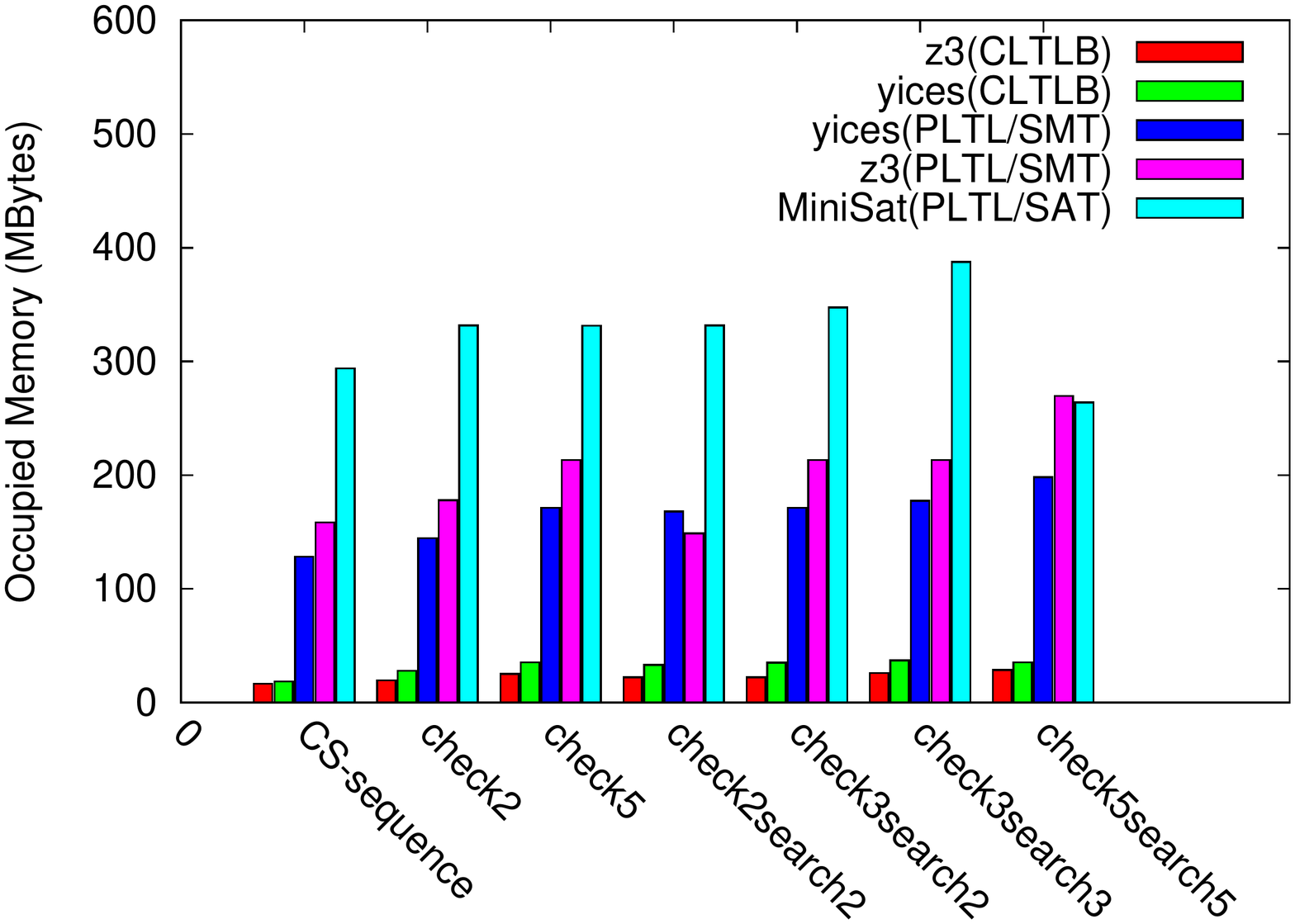}
   \label{fig:realcaseM}
 }
\subfigure[Elapsed times on the third set of experiments]{
   \includegraphics[width=.9\columnwidth]{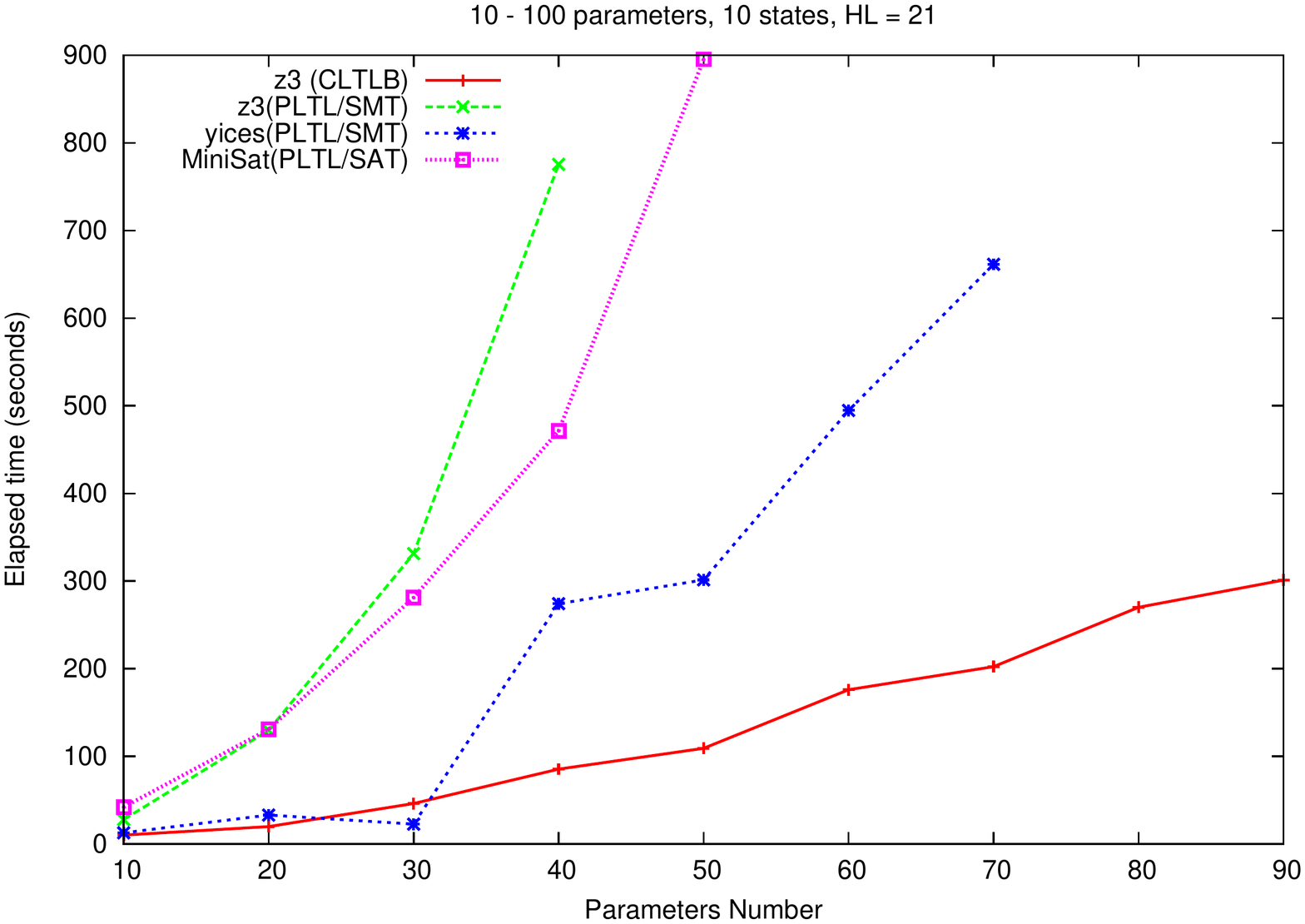}
   \label{fig:abscaseT}
 }
 \subfigure[Memory occupations on the third set of experiments]{
   \includegraphics[width=.9\columnwidth]{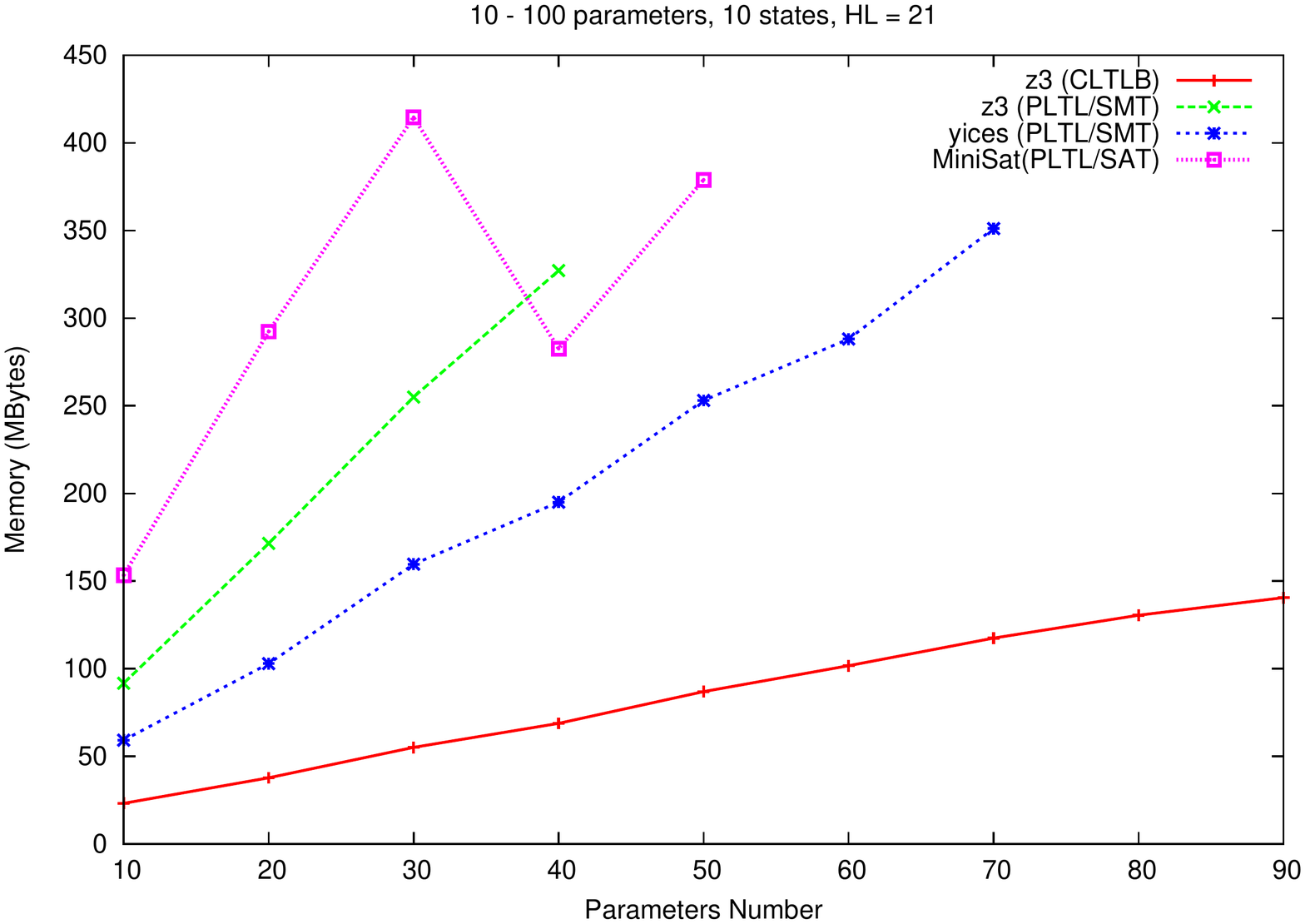}
   \label{fig:abscaseM}
 }

 \caption{Experimental Results}
 \label{fig:performances}
\end{figure*}

In order to evaluate the encoding presented in this paper we built a plug-in of
Zot and we used it in three sets of experiments\footnote{The experiments sets are available at \\ http://home.dei.polimi.it/cavallaro/sefm10-experiments.html}:
(i) We created adapters for sequences of increasing length related to the case study presented in Section~\ref{sec:casestudy}. This set of experiments was used as a qualitative evaluation of the approach on examples taken from the real world.
(ii) We ran the same set of experiments on Zot using three different encodings, namely the traditional SAT-based encoding (PLTL/SAT), the new SMT-based one of the same logic (PLTL/SMT), and the SMT-based of logic CLTLB(DL) introduced in this paper. We measured elapsed time and occupied memory, and we compared the results to get an estimate of how the introduction of the SMT-solver speeds up the adapter-building mechanism.
(iii) We created some service interface models with growing number of parameters
and tried to solve them with both the original version of the encoding and with the extensions.
This set of experiments has the purpose to compare how much the new encoding scales on models larger than those found in common practice.

All experiments were run using the Common Lisp compiler SBCL 1.0.29.11 on a 2.50GHz Core2 Duo laptop with Linux and 4 GB RAM.
We chose to use two different SMT-solvers in our tests: Microsoft
Z3\footnote{Z3: http://research.microsoft.com/en-us/um/redmond/projects/z3/} and SRI Yices\footnote{Yices: http://yices.csl.sri.com/}.
For the SAT-based PLTL
encoding we used MiniSat\footnote{MiniSat: http://minisat.se/}.

The first set of experiments was carried out selecting some operation sequences
on the expected service presented in Section~\ref{sec:casestudy}. The
selected sequences set comprises the simple sequence analyzed in the case study
plus sequences of growing length obtained trying to execute up to 5 consecutive
\verb+searchSongs+ and \verb+checkSongExists+ operations.
We set the time bounds for the experiments using a simple heuristic, based on the sum of the states of the automata of the input services.
In those cases in which the abstract sequence
featured repeated invocations of the same operation, the time bound was augmented with the number of repetitions of each operation.
This set of experiments produced a set of mapping scripts that we checked by inspection. 
%
Fig. \ref{fig:realcaseT} and Fig. \ref{fig:realcaseM} report the overall results. Fig. \ref{fig:realcaseM} shows that the CLTLB(DL) encoding has lower memory occupation than the SAT-based PLTL encoding for the same problem. Fig. \ref{fig:realcaseT} shows that the CLTLB(DL) encoding on Z3 performs much better than the others.

Lastly,
we tried to push the limits of our technique to check its robustness. 
To do so, we generated simple service protocols featuring operations with a
growing number of parameters. We chose this setting for our experiments based on
our experience in the common practice, which suggests that services
usually exhibit very simple protocols, while operations have sometimes a considerable number of parameters. 
Note that the models used in these experiments are much bigger than those commonly found in practice. The experiments are based on
expected and actual services with 10 states, and a trace
bound of 21 time instants.
The results are shown in Fig. \ref{fig:abscaseT} and in Fig. \ref{fig:abscaseM}. 
The number of parameters used in experiments
ranges from 10 (i.e. each operation has 10 input and 10 output
parameters) to 90. As shown in the figures, the CLTLB(DL) encoding on Z3
was the only one we managed to push up to 90 parameters, while we
stopped experimenting much earlier with the PLTL encoding on Yices, Z3
and MiniSat.
Note that in Fig. \ref{fig:abscaseT}-\ref{fig:abscaseM} the combination CLTLB(DL)/Yices is missing because of its poor performance on this set of experiments (the simplest case was solved in more than 500 seconds). 

\section{Related Work} \label{sec:relatedwork}

Our approach is closely related both to works supporting substitution of services and to works about verification using model checking.
Many approaches that support the automatic generation of adapters
(or equivalent mechanisms) are based on the use of ontologies and
focus on non-conversational services (see for instance
\cite{cavallaro-pesos-09, drumm-2008}).
They all assume that the usual WSDL definition of a service interface is
enriched with some kinds of ontological annotations. At run-time,
when a service bound to a composition needs to be substituted, a
software agent generates a mapping by parsing such ontological
annotations.
\textit{SCIROCO}~\cite{1439100} offers similar features but focuses on stateful services. It requires all services to be
annotated with both a SAWSDL description and a WS\-ResourceProperties~\cite{WS-ResourceProperties} document, which
represents the state of the service. When an invoked service becomes
unavailable, \textit{SCIROCO} exploits the SAWSDL annotations to
find a set of candidates that expose a semantically matching
interface. Then, the WS-ResourceProperties document associated to
each candidate service is analyzed to find out if it is possible to
bring the candidate in a state that is compatible with the state of
the unavailable service.
If this is possible, then this service is selected for replacement of the one that is unavailable. All
these three approaches offer full run-time automation for service
substitution, but as the services they consider are not
conversational, they perform the mapping on a per-operation basis.
An approach that generates adapters covering the case of interaction protocols
mismatches is presented
in~\cite{DBLP:conf/icsoc/BrogiP06}.
It assumes to start from a service composition and a service behavioral description both
written in the BPEL language~\cite{BPEL}.
These are then translated in the \emph{YAWL} formal language~\cite{DBLP:journals/is/AalstH05}
and matched in order to identify an invocation trace in the service behavioral description that matches the one expected by the service
composition. The matching algorithm is based on graph exploration and considers both control flow and data flow requirements.
The approach presented in~\cite{contract-generation-foclasa} offers similar features and has been implemented in an open source
tool\footnote{{The Dinapter tool: {\scriptsize http://sourceforge.net/projects/dinapter}}}.
While both these approaches appear to fulfill our need for supporting interaction protocol mapping, they present
some shortcoming in terms of performances, as shown in~\cite{CNP09}.

Although QF-UFIDL involves variables over infinite domain, our
particular BMC of CLTLB(DL) formulae became effective because it is
not used as an infinite-state model checking procedure.
In general, transitions systems defined by arithmetic constraints provide a large class of
infinite-state systems which are suitable for modeling a large variety of
applications.
So, intensive work has been devoted to identify useful classes with
decidable reachability and safety properties \cite{FO97, CJ98}.
Some implemented procedures \cite{B98, TReX}
rely on a pure operational approach and the
complexity of the decision problem of the underlying arithmetic
(3-EXPTIME in the case of Presburger Logic) do not make them
appropriate for runtime checking.
Much effort is also devoted to study decidabilty and complexity of
temporal logic of arithmetic constraints, \cite{D04, DD02, BFMPRS10, CC00}.
\cite{DFGD06} proposes a semi-decision procedure aimed to be
used for model checking of an extension of CTL* with Presburger
constraints.
Finally, an operational approach to BMC which exploits a direct
translation of LTL formulae of arithmetic constraints is suggested in \cite{dMRS02}.
Our approach offers a mixed operational-descriptive BMC based
on the satisfiability of CLTLB(DL) formulae which enjoys the
NP-completeness of the decision problem of DL, significantly less than that of
more complex theories.

\section{Conclusion}\label{sec:conclusion}
In this paper we introduced an efficient encoding for a linear temporal logic with arithmetic constraints.
Our encoding was found very suitable for application to a real problem taken from the SOA domain and showed better performances and lower memory occupation than the other encodings we compared it with.
The research work is currently still ongoing. For future work we plan to further experiment with our encoding and to investigate its theoretical properties.
\section*{Acknowledgments}
Many thanks to
Elisabetta Di Nitto, Angelo Morzenti, and
Pierluigi San Pietro for the fruitful discussions and their support.
This research has been partially funded by the European Commission,
Programme IDEAS-ERC, Project 227977-SMScom, and by the Italian
Government under the project PRIN 2007 D-ASAP (2007XKEHFA).

\bibliographystyle{IEEEtran}
\bibliography{sefm10-soa-bib}

\end{document}